\documentclass[fleqn,twoside]{article}
\usepackage{espcrc2}

\usepackage{graphicx}
\usepackage[figuresright]{rotating}

\newcommand{\AmS}{{\protect\the\textfont2
  A\kern-.1667em\lower.5ex\hbox{M}\kern-.125emS}}

\title{
{
\vspace{-3.0cm} \normalsize \hfill
\parbox{30mm}{DESY 02-138}
}\\[15mm]
Speeding up the HMC: QCD with Clover-Improved Wilson Fermions}

\author{M. Hasenbusch \address[NIC]{NIC/DESY-Zeuthen, 
        Platanenallee 6, D-15738 Zeuthen, Germany}%
        and
        K. Jansen\addressmark[NIC]}
       
\begin{document}

\begin{abstract}
We apply a recent proposal to speed up the Hybrid-Monte-Carlo simulation
of systems with dynamical fermions to two flavor QCD with
clover-improvement. For our smallest quark masses we see a speed-up of more
than a factor of two compared with the standard algorithm.
\vspace{1pc}
\end{abstract}

\maketitle

\section{Introduction}

It is clear that simulation algorithms that are used today
will not be able to reach physical values of the quark masses. 
The scaling
behavior (see e.g. \cite{Lippert})
of the algorithms predict enormous costs for simulations
at quark masses as light as the up- and down-quarks. To reach this
physical point, extrapolations using $\chi$PT (chiral perturbation theory)
have to be used. However contact to $\chi$PT seems to be happening at
rather small values of the quark masses themselves.
Any progress to render simulations easier, when approaching the small
quark mass regime will help therefore to reach overlap between
$\chi$PT and lattice QCD, allowing for a safe extrapolation to
physical quark masses.

Still the HMC (Hybrid-Monte-Carlo) algorithm \cite{hybrid}
and its variants are the methods of choice
in large scale simulations of lattice QCD with dynamical Wilson-fermions.
One obstacle in going to light quarks is that the step size of the integration 
scheme has to be reduced to maintain a constant acceptance rate.
In ref. \cite{MH_schwinger} we proposed to  split the fermion matrix into two
factors and to introduce a pseudo-fermion field for both factors.
The numerical study of the two dimensional Schwinger model showed that 
the step-size can be enlarged and thus the 
computational effort can be reduced substantially this way.
In ref. \cite{ourlattice2001} we presented first results for lattice 
QCD with two flavors of Wilson fermions and clover improvement \cite{clover}. 
Here we present new results for lattices up to $16^3 \times 24$ at 
$\beta=5.2$  and quark masses down to $m_{PS}/m_{V} \approx 0.686$.

\section{The pseudo-fermion action}
The partition function of lattice QCD with two degenerate 
flavors of dynamical fermions is given by
\begin{equation}
\label{partition}
 Z = \int \mbox{D}[U] \exp(-S_G[U])  \;\; \mbox{det} M[U]^2 \;\;,
 \end{equation}
where $S_G[U]$ is the Wilson plaquette action. In our case, 
$M[U]$ is the Wilson fermion  
matrix with $O(a)$ (clover)-improvement. For the details see e.g. 
ref. \cite{JansenLiu}.
An important feature of our proposal \cite{MH_schwinger} is that it can be
applied on top of standard preconditioning.
Here, 
we use even-odd preconditioning as it is detailed in ref. \cite{JansenLiu}.
The fermion matrix can be written as
\begin{equation}
 M=  \left( \begin{array}{cc}
    1_{ee}+T_{ee}   &  -\kappa M_{eo}  \\
    -\kappa M_{oe}   &  1_{oo}+T_{oo}
\end{array} \right)  \; \; \;,
\end{equation}
where $e$ refers to even sites and $o$ to odd sites of the lattice.
The determinant of the fermion matrix can now be written as
\begin{equation}
\label{ourevenodd}
\mbox{det} M \propto \mbox{det} (1_{ee} + T_{ee}) \; \mbox{det} \hat M
\;\;,
\end{equation}
where
$
\hat M =
1_{oo} + T_{oo} - M_{oe} (1_{ee} + T_{ee})^{-1} M_{eo}
$.
In the following, we shall consider the Hermitian matrix
$
 \hat Q =  \hat c_0 \gamma_5 \hat M \;$,
where we have set $\hat c_0 = 1$ in the simulations discussed below.
The effective action for the standard HMC simulation 
reads \cite{JansenLiu}
\begin{equation}
 S_{eff}[U,\phi^{\dag},\phi] =
 S_G[U] + S_{det} [U] + S_F[U,\phi^{\dag},\phi] \;,
\end{equation}
with
$S_{det}[U]   =  -2 \mbox{Tr} \log(1+T_{ee})$ 
and $S_F[U,\phi^{\dag},\phi]   = \phi^{\dag} \hat Q^{-2} \phi$ .
In our study we keep $S_G[U]$ and $S_{det}[U]$ in their standard form.
However,
$S_F[U,\phi^{\dag},\phi]$ is replaced by alternative expressions: 
We split the fermion matrix $\hat Q$ into two factors.
The determinant of each factor is estimated by an integral over
pseudo-fermion fields:
\begin{equation}
\label{general}
 \mbox{det} \hat Q^2  =
 \mbox{det} W  W^{\dag}\; \mbox{det} [W^{-1} \hat Q]
 [W^{-1} \hat Q]^{\dag} 
 \propto  
\end{equation}
\vskip-0.4cm
\begin{eqnarray}
 \int \mbox{D} \phi_1^{\dag} \int \mbox{D} \phi_1 
 \int \mbox{D} \phi_2^{\dag} \int \mbox{D} \phi_2 
     \exp\left(
      -S_{F1}  - S_{F2}   
   \right)  \nonumber 
\end{eqnarray}
where
\begin{eqnarray}
S_{F1} &=&
\phi_1^{\dag} \left(W W^{\dag}\right)^{-1} \phi_1\;\;\; , \nonumber \\
S_{F2} &=& \phi_2^{\dag} \left([W^{-1} \hat Q]
         [W^{-1} \hat Q]^{\dag}\right)^{-1} \phi_2 \;\;\;.
\end{eqnarray}
We considered two choices for $W$:
\begin{equation}
\label{orginaltilde}
W = \hat Q + \rho \gamma_5 \;\;,
\end{equation}
which is the original proposal of ref. \cite{MH_schwinger}. Below we shall 
show only results for this choice.
In addition we considered
$W = \hat Q + i \rho $,
which was first tested in ref. \cite{ourlattice2001}. Also in 
ref. \cite{Knechtli},
results for this choice are presented. It turns out that
both choices for $W$ give a similar performance improvement of the algorithm.

An important feature of our approach is that
the variation of the modified pseudo-fermion action can be computed as easily
as in the standard case.
The variation of $S_{F1}$ is essentially the same as 
for the standard pseudo-fermion action.
One only has to replace $\hat Q$ by $W$.
For the second part of the pseudo-fermion action we get
\begin{eqnarray}
\label{seconddiff}
\delta S_{F2} = &-& X^{\dag} \; \delta \hat Q \;   Y  \;\;\;
 - \;\; Y^{\dag} \; \delta \hat Q^{\dag} \; X \nonumber \\
  &+& X^{\dag} \; \delta W \; \phi_2 \;\;
  + \; \phi_2^{\dag} \; \delta W^{\dag} \; X
  \end{eqnarray}
  with the vectors
  \begin{equation}
 \label{XYgeneral2}
 X = \left(\hat Q \hat Q^{\dag} \right)^{-1} \;W\; \phi_2  \;\;,
 \;\; Y = \hat Q^{-1} \;W \; \phi_2 \;\;\;.
\end{equation}

\section{Numerical results}
We have tested our modified algorithm at parameters that had been
studied by UKQCD before \cite{Sroczynski}. We have performed simulations
at $\beta=5.2$ and $c_{SW} = 1.76$. Note that $c_{SW} = 1.76$ was a
preliminary result for the improvement coefficient, while the final
analysis resulted in $c_{SW} = 2.0171$ for $\beta=5.2$ \cite{alphacsw3}.
We have studied $\kappa=0.137$, $0.139$, $0.1395$ and $0.1398$. These
values of $\kappa$ correspond to $m_{\pi}/m_{\rho}\approx0.856$, $0.792$,
$0.715$ and $0.686$, respectively \cite{Sroczynski}.
We applied periodic boundary conditions in all lattice directions,
except for anti-periodic boundary conditions in time-direction
for the fermion-fields.

In addition to the standard leap-frog scheme, we used a 
scheme with a reduced  coefficient of the 
$O(\delta \tau^2)$ corrections proposed by Sexton and 
Weingarten (see eq.~(6.4) of ref. \cite{SeWe}). 
In order to eliminate the influence of the gauge action on the step size of 
the integration scheme, we used the split of time scale as proposed in
ref. \cite{SeWe}. In particular, we have computed the variation of the 
gauge action four times as frequent as for the fermion action. 
As length of the trajectory we have always chosen $\tau=1$.

In our study, we applied the BiCGstab as solver. We have stopped the 
solver when the iterated residual $r$ becomes smaller than a certain bound.
To compute the action for the accept/reject step at the end of the trajectory
we required $r^2 < 10^{-20}$. Note that we take the absolute  residual and 
not the relative.
To compute the variation of the fermionic action, we required a less 
strict criterion $r^2 < R^2$.  It turns out that for $R^2$ smaller than some
threshold the acceptance rate virtually does not depend on $R^2$, while 
it rapidly drops to zero as $R^2$ becomes larger than this threshold.
In the simulations reported below, $R^2$ is chosen 
$10^{-1}$ times the threshold or smaller.


In table \ref{tableacc} we show results for the acceptance rate for a 
$8^3 \times 24$ lattice at $\kappa=0.137$. After equilibration we generated
200 trajectories for each parameter set. The standard pseudo-fermion action 
is given by $\rho=0$. First of all, for the same step size $\delta \tau$
the acceptance with the modified pseudo-fermion action is higher than 
for the standard action.
The maximum of the acceptance rate is rather shallow for both integration 
schemes.  It is located at $\rho \approx 0.5$.
\begin{table}
\caption{\sl \label{tableacc}
Acceptance rates $P_{acc}$  for the
$8^3 \times 24$ lattice at
$\beta=5.2$, $\kappa= 0.137$ and
$c_{sw}=1.76$.
Each run consists of 200 trajectories.
The leap-frog runs (L) were performed with the step size $\delta \tau = 0.02$.
The runs with the improved scheme (S) with $\delta \tau = 0.05$.
}
\vskip0.2cm
\begin{center}
\begin{tabular}{cccl}
\hline
$\rho$ & L,$P_{acc}$ & S,$P_{acc}$ \\
\hline
0.0  &  0.856(7) & 0.876(10) \\
0.1  &  0.914(8) &  \\
0.3  &  0.940(5) & 0.968(4)\\
0.5  &  0.948(5) & 0.973(2)\\
0.6  &  0.944(4) & 0.973(2)\\
0.7  &  0.934(4) & 0.969(3)\\
1.0  &  0.938(7) & 0.957(3)\\
\hline
\end{tabular}
\end{center}
\vskip-1.1cm
\end{table}
Next we performed longer simulations (with 6000 to 8000 trajectories each)
for the standard pseudo-fermion action
($\rho=0$) and the optimal $\rho=0.5$ in the modified pseudo-fermion case.
This time we tuned the step-size such that
$P_{acc} \approx 0.8$. For the standard pseudo-fermion action and the leap-frog
scheme we get with $\delta \tau = 0.025$ the acceptance rate
$P_{acc} = 0.793(3)$. The leap-frog scheme with 
the modified pseudo-fermion
action at $\rho=0.5$ 
gives $P_{acc} = 0.770(3)$ for  $\delta \tau = 0.04$. 
Using the improved integration scheme, 
combined with the modified pseudo-fermion action
even  a step-size as large as $\delta \tau = 0.1$ 
gives $P_{acc} = 0.883(2)$.
Note however that for the improved scheme the variation of the 
action has to be computed twice per time-step and for the leap-frog only once.
For these more extended runs we have computed integrated autocorrelation 
times for the value of the plaquette and the number of iterations that are 
needed by the solver. It turned out that within error-bars the autocorrelation
times are the some for all three runs reported above. Hence the modification of 
the pseudo-fermion action has no detectable effect on autocorrelation times
(measured in units of trajectories).

In table \ref{many} we give our results  for the $16^3 \times 24$ lattice at 
our largest value of $\kappa$. 
Starting after equilibration, we performed 100 trajectories for each parameter
set. Here the maximum of the acceptance rate has become sharper than  for 
the $8^3 \times 24$ lattice at $\kappa=0.137$. For the leap-frog scheme the 
largest acceptance rate is still reached at $\rho \approx 0.5$, 
while for the scheme
of Sexton and Weingarten the optimum is shifted down to $\rho \approx 0.2$. 
It is interesting to note that the improved scheme profits much 
more from the modified pseudo-fermion action than the leap-frog scheme.
In the case of the leap-frog scheme the step-size can only be doubled, while 
for the improved scheme the step-size can be more than tripled. In the last
column of table 2 we give the total number of applications of $\hat Q^2$
per trajectory. This number is proportional to the numerical effort of the 
simulation.  Comparing the leap-frog scheme at $\rho=0$ with the improved 
scheme at $\rho=0.2$, we find a reduction of the numerical cost by a factor
of roughly $2.4$.

\begin{table}
\caption{\sl \label{many}
Results for
a $16^3 \times 24$ lattice at $\beta=5.2$, $c_{sw}=1.76$
and $\kappa=0.1398$. 
$\# \hat Q^2$ gives the total number of applications of  $\hat Q^2$
per trajectory.}
\vskip0.2cm
 \begin{center}
\begin{tabular}{clllr}
\hline
scheme & $\rho$ & $\delta \tau $ &  $P_{acc}$ & $\# \hat Q^2$ \\
\hline
  L    & 0.    & 0.01 &  0.77(3)   & 17000\\
  L    & 0.05  & 0.02 &  0.62(3)   & 11700\\
  L    & 0.15  & 0.02 &  0.75(3)   & 10700\\
  L    & 0.3   & 0.02 &  0.76(2)   & 10100\\
  L    & 0.5   & 0.02 &  0.78(2)   &  9500\\
  L    & 1.0   & 0.02 & 0.64(4)    &  9300\\
\hline
  S    & 0.    & 0.02 & 0.64(4)    &  16500\\
  S    & 0.05  & 0.066.. & 0.35(4) & 7500 \\
  S    & 0.1   & 0.066.. & 0.67(3) & 6500\\
  S    & 0.15  & 0.066.. & 0.72(3) & 6800\\
  S    & 0.2   & 0.066.. & 0.74(3) & 6900\\
  S    & 0.4   & 0.066.. & 0.49(3) & 6100\\
  S    & 0.5   & 0.05    & 0.67(4) & 7700 \\
\hline
\end{tabular}
\end{center}
\vskip-0.8cm
\end{table}

\end{document}